\documentclass[english,aps,prl,twocolumn,superscriptaddress,showpacs,reprint]{revtex4-1}
\usepackage{lmodern}
\usepackage[T1]{fontenc}
\usepackage[latin9]{inputenc}
\usepackage{xcolor}
\usepackage{pdfcolmk}
\usepackage{babel}
\usepackage{textcomp}
\usepackage{amssymb}
\usepackage{graphicx}
\PassOptionsToPackage{normalem}{ulem}
\usepackage{ulem}
\usepackage{subscript}
\usepackage[unicode=true,pdfusetitle,
 bookmarks=true,bookmarksnumbered=false,bookmarksopen=false,
 breaklinks=false,pdfborder={0 0 0},backref=false,colorlinks=true]
 {hyperref}

\begin{document}

\title{Electronic structure basis for the titanic magnetoresistance in WTe\textsubscript{2}}

\author{I. Pletikosi\'{c}}

\affiliation{Department of Physics, Princeton University, Princeton, New Jersey
08544, USA}

\affiliation{Condensed Matter Physics and Materials Science Department, Brookhaven
National Laboratory, Upton, New York 11973, USA}

\author{Mazhar N. Ali}

\affiliation{Department of Chemistry, Princeton University, Princeton, New Jersey
08544, USA}

\author{A. Fedorov}

\affiliation{Advanced Light Source, Lawrence Berkeley National Laboratory, Berkeley,
California 94720, USA}

\author{R. J. Cava}

\affiliation{Department of Chemistry, Princeton University, Princeton, New Jersey
08544, USA}

\author{T. Valla}

\affiliation{Condensed Matter Physics and Materials Science Department, Brookhaven
National Laboratory, Upton, New York 11973, USA}
\begin{abstract}
The electronic structure basis of the extremely large magnetoresistance
in layered non-magnetic tungsten ditelluride has been investigated
by angle-resolved photoelectron spectroscopy. Hole and electron pockets
of approximately the same size were found at the Fermi level, suggesting
that carrier compensation should be considered the primary source
of the effect. The material exhibits a highly anisotropic, quasi one-dimensional
Fermi surface from which the pronounced anisotropy of the magnetoresistance
follows. A change in the Fermi surface with temperature was found
and a high-density-of-states band that may take over conduction at
higher temperatures and cause the observed turn-on behavior of the
magnetoresistance in WTe\textsubscript{2} was identified. 
\end{abstract}


\maketitle
An enormous (``titanic'') positive magnetoresistance  was recently
discovered in several non-magnetic materials including Cd\textsubscript{3}As\textsubscript{2},
WTe\textsubscript{2}, and NbSb\textsubscript{2} \cite{Liang2014,Ali2014,Cedo2014}.
For each of these an increase in resistivity of several orders of
magnitude was found at low temperatures, with no evidence of saturation
in very high fields. This is in strong contrast with the conventional
behavior observed in simple metals where the magnetoresistance is
usually very small, quadratic in low fields, and tends to saturate
in high fields \cite{Pippard1989}. The observed \emph{titanic} effect
is reminiscent of that seen in pure bismuth, where a large magnetoresistance
stems from small effective masses, long mean free paths and small
and balanced concentrations of holes and electrons \cite{Yang1999,Yang2000}.
To understand the \emph{titanic} magnetoresistance, it would be highly
desirable to see what attributes of the electronic structure are at
play in these materials. In at least one of them (Cd\textsubscript{3}As\textsubscript{2}),
the observed effect has been related to the three-dimensional (3D)
Dirac nature of its electronic structure, where breaking of time-reversal
symmetry in magnetic field removes the topological protection and
opens the backscattering channels, leading to a large increase of
resistivity. For the other two materials, however, the calculated
Fermi surfaces and the overall electronic band-structure do not show
a resemblance to a 3D Dirac semimetal \cite{Ali2014,Cedo2014}. As
a matter of fact, both materials have complicated band-structure with
multiple Fermi surfaces and display a very anisotropic magnetoresistance,
rapidly disappearing with increasing temperature, suggesting that
the fine details of the electronic structure play a significant role. 

Our study focuses on WTe\textsubscript{2}, a layered transition metal
dichalcogenide in which each tungsten (transition metal) layer is
surrounded by two tellurium (chalcogen) layers. Owing to pronounced
bonding between W atoms (bond length is only about 4\% larger than
in tungsten metal), the layers are structurally distorted from the
usual hexagonal network: tungsten atoms form slightly buckled, well
separated zig-zag chains, and the tellurium octahedra surrounding
each W atom distort to accommodate them \cite{Augustin2000,Ali2014}.
The layers are held together by van der Waals interaction, and are
sequentially rotated by 180\textdegree{} to account for the layer
buckling.

The temperature dependence of resistivity, Hall coefficient and thermoelectric
power have been successfully explained for WTe\textsubscript{2} by
a three-carrier model \cite{Kabashima1966}. The three bands, however,
were not clearly identified, due to the complex band structure of
WTe\textsubscript{2}, with many bands interwoven along the direction
of the tungsten chains as shown in a band structure density functional
study by Augustin \emph{et al}. \cite{Augustin2000}. The experimental,
angle-resolved photoemission spectroscopy (ARPES) results from the
same study \cite{Augustin2000}, were not of sufficiently high angular
and energy resolution to get any conclusive insight into the states
responsible for conduction in the vicinity of the Fermi level.

Here we investigate the low energy electronic structure of WTe\textsubscript{2}
using high-resolution ARPES. We find that the configuration and the
temperature dependence of its Fermi surface, as well as band dispersion
along the direction coinciding with the direction of the metallic
chains can explain the main features of the \emph{titanic} magnetoresistive
effect. Our results suggest that the large unidirectional magnetoresistance
in WTe\textsubscript{2} originates from small and, at low T, perfectly
balanced electron and hole Fermi pockets (tubes) that almost touch
and are aligned along the tungsten chain direction. The effect disappears
as the balance shifts and the contribution from a flat shallow band
increases with temperature.

Single crystal platelets of WTe\textsubscript{2} were grown via bromine
vapor transport. Purified tellurium and tungsten powders were ground
together and sintered in an evacuated quartz tube for two days at
700\textdegree C and two days at 750\textdegree C, with a homogenization
step in between. The resulting polycrystalline pellet was finely ground,
sealed in a tube with \textasciitilde{}3 mg/ml of bromine, and then
heated in a three-zone furnace with a 100\textdegree C temperature
gradient applied between 750\textdegree C and 650\textdegree C for
a week \cite{Ali2014}.

The ARPES measurements were conducted using a Scienta SES2002 analyzer
at U13, Scienta R4000 at the U5 beam line of the National Synchrotron
Light Source at BNL ($h\nu=21\,\mathrm{eV}$), and at the 12.0.1 beam
line of the Advanced Light Source at LBNL (38--78~eV) using a Scienta
SES100 analyzer. The total experimental resolution was \textasciitilde{}15~meV
and <0.2\textdegree{} in all three experimental setups. The two-dimensional
Brillouin zone mapping was accomplished by sample rotation perpendicularly
to the analyzer slit, in steps of 2\textdegree , 0.5\textdegree ,
or 0.25\textdegree . The samples were glued to the holder by a conductive
epoxy resin and cleaved in ultrahigh vacuum ($p<10^{-8}~\mathrm{Pa}$)
just before the measurements. Sample cooling was provided through
contact with cryostats filled with liquid helium or liquid nitrogen.

\begin{figure}
\includegraphics[width=86mm]{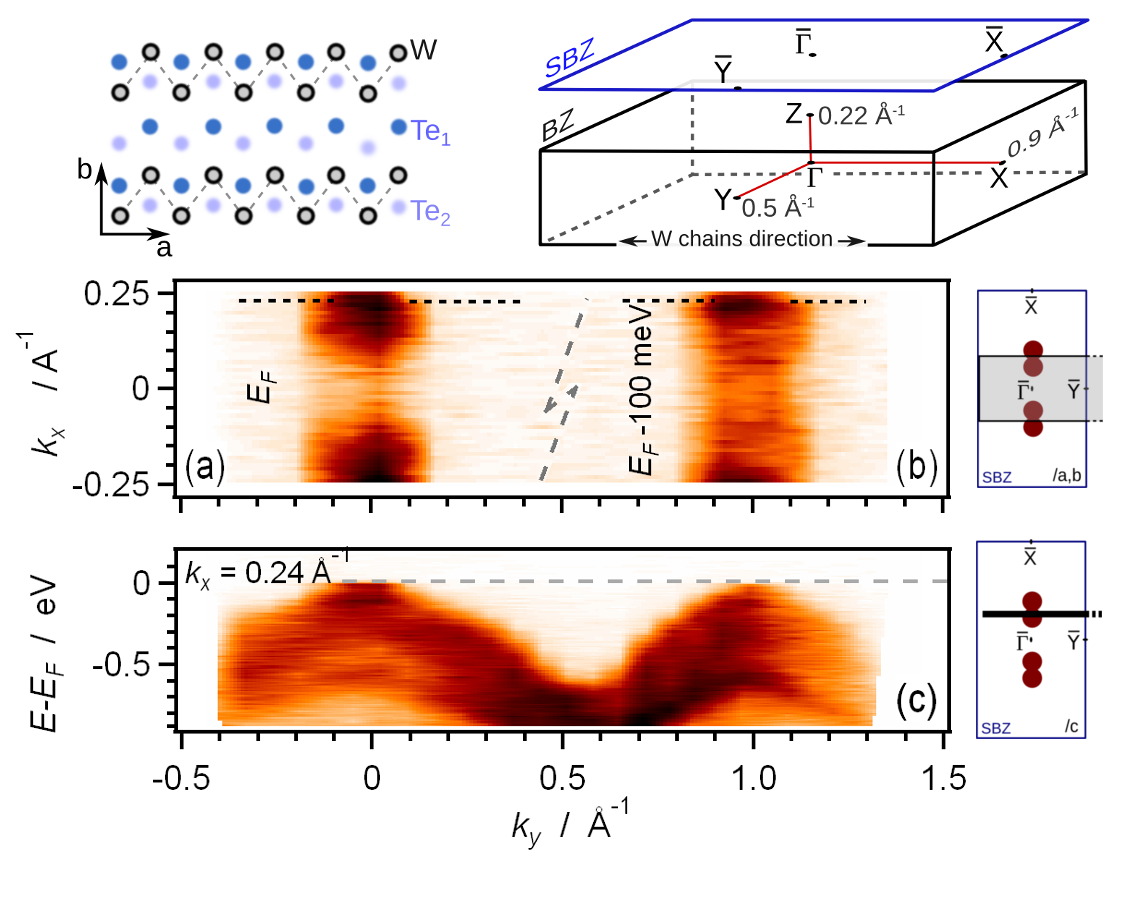}

\protect\caption{\label{fig:twoSBZ}Electronic structure of tungsten ditelluride mapped
by ARPES through two extended surface Brillouin zones along the $\bar{\Gamma}-\bar{Y}$
($k_{y}$) direction. Constant energy maps were extracted (a) at the
Fermi level, $E_{F}$, and 100~meV below (b). Only a portion of the
zone that extends to $\pm0.9\,\protect\AA^{-1}$ was mapped along
the $\bar{\Gamma}-\bar{X}$ ($k_{x}$) direction. (c) Band dispersion
along $\bar{\Gamma}-\bar{Y}$ at the point $k_{x}=0.24\,\protect\AA^{-1}$
with the highest intensity at the Fermi level. Excitation energy $h\nu=21\,\mathrm{eV}$;
sampling around the perpendicular momentum of $k_{z}=2.7\,\protect\AA^{-1}$
($\Gamma_{6}$). The map was assembled from a series of spectra acquired
with 2\textdegree{} polar rotations. The sample temperature was 20~K.
Top: Schematic view of the crystal structure, and bulk (BZ) and surface
(SBZ) Brillouin zones. Right: a reference sketch of the SBZ with ARPES
scans performed in (a)--(c).}
\end{figure}

Structural reduction to well-separated WTe\textsubscript{2} layers,
and within each layer to chains of tellurium-surrounded tungsten atoms
leaves a substantial mark in the low-energy band structure. The constant
energy maps at the Fermi level and 100~meV below, shown in Fig. \ref{fig:twoSBZ},
exhibit pronounced unidirectionality, in the sense that the charge
carriers acquire only a small momentum perpendicular to the chains
($k_{y}$ of about $0.07\,\AA^{-1}$), and several times larger along
the chains.   The impurity scattering pattern, which is, to the
first approximation, an autocorrelation map of the Fermi surface,
will inherit this directionality, meaning that the charge carriers
will preferably scatter along the chains, regardless of the current
direction. This may thus lead to the reported anisotropy of the magnetoresistance
\cite{Ali2014}.

\begin{figure}
\includegraphics[width=50mm]{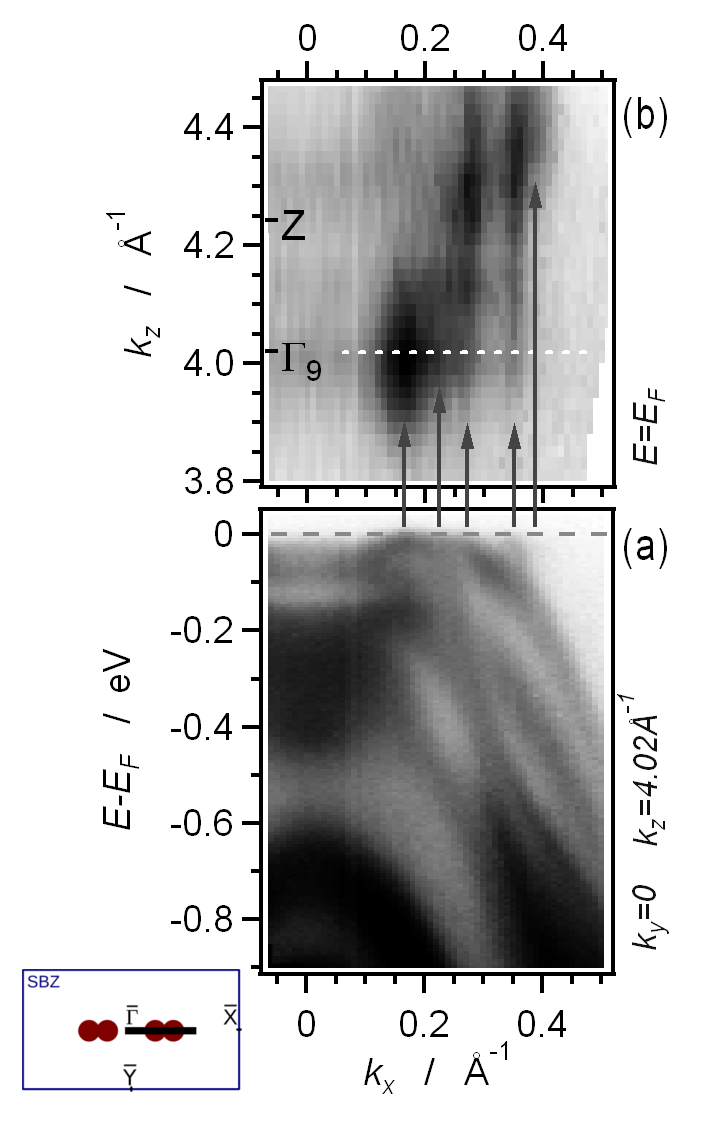}

\protect\caption{\label{fig:kz}(a) Dispersion of WTe\protect\textsubscript{2} bands
along the $\bar{\Gamma}-\bar{X}$ direction taken with the excitation
energy of $h\nu=54\,\mathrm{eV}$. The $k_{z}$ dependence of the
states at the Fermi level, $E_{F}$, shown in (b), was obtained by
combining many of such spectra taken in the 40--70~eV range of excitation
energies. The band structure in (a) represents the cut at $k_{z}=4.02\,\protect\AA^{-1}$,
which is the center of the ninth extended Brillouin zone, $\Gamma_{9}$.}
\end{figure}

Figure \ref{fig:twoSBZ}(c) shows that the band dispersion in the
perpendicular direction (along the c-axis) is modest, amounting to
some 0.6~eV over the whole Brillouin zone. Even though Fig. \ref{fig:twoSBZ}(a)
shows only a part of the Brillouin zone, which extends to $\pm0.90\,\AA^{-1}$
in the chain direction (due to limited angular window of the ARPES
scan), further scanning found no bands close to the Fermi level except
those along $\bar{\Gamma}-\bar{X}$. One such scan is shown in Fig.
\ref{fig:kz}(a). The only states crossing the Fermi level are found
for $k_{x}$ between $0.15\,\AA^{-1}$ and $0.40\,\AA^{-1}$, in agreement
with density functional predictions in Refs. \onlinecite{Augustin2000}
and \onlinecite{Ali2014}. 

\begin{figure}
\includegraphics[width=86mm]{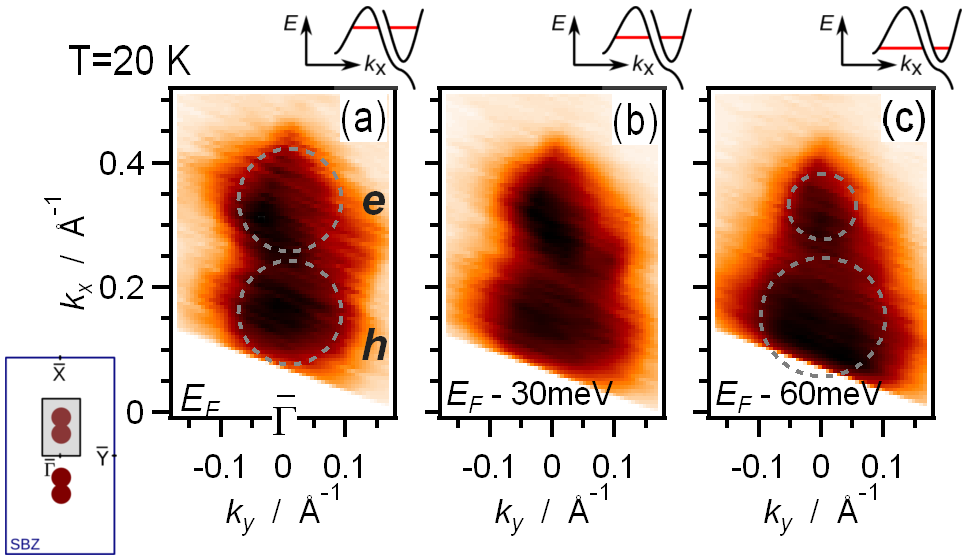}

\protect\caption{\label{fig:FS20K}(a)--(c) Constant energy contours at T=20~K in
the region of the Brillouin zone where the bands of WTe\protect\textsubscript{2}
cross the Fermi level, $E_{F}$. Electron-like and hole-like pockets
are marked with \emph{e} and \emph{h}, respectively. Excitation energy
was $h\nu=21\,\mathrm{eV}$. The mapping has been carried out by 0.5\textdegree{}
polar rotation. Dashed circles have been added as a guide to the eye.
Above: a sketch of the band structure and constant energy cuts being
made in (a)--(c).}
\end{figure}

\begin{figure*}
\includegraphics[width=180mm]{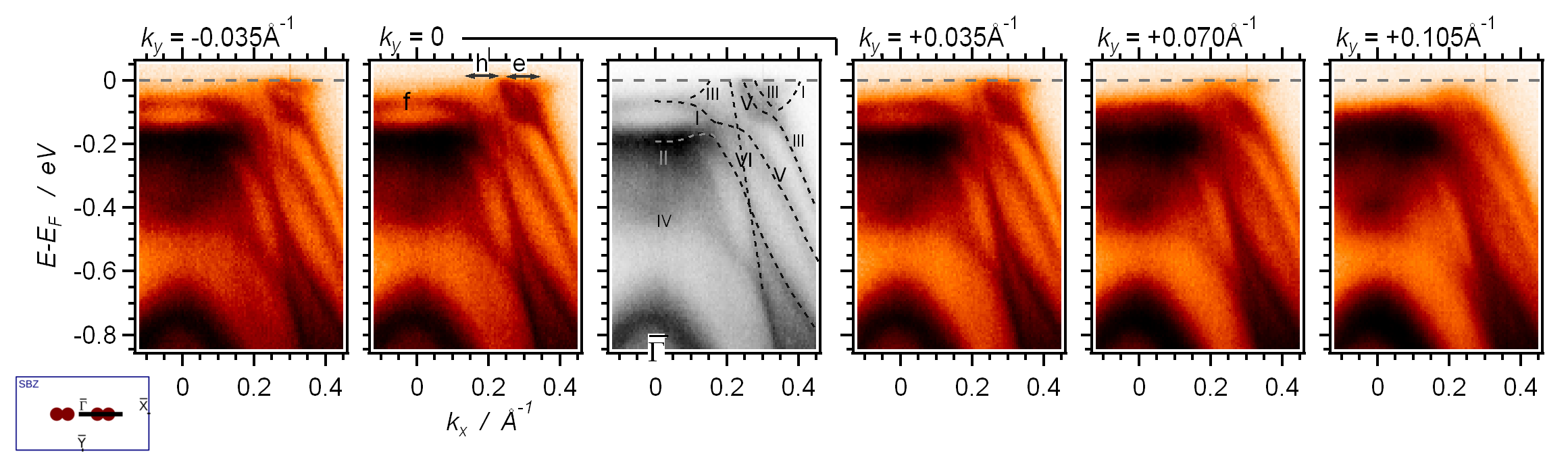}

\protect\caption{\label{fig:Dispersion_maps}High resolution band dispersion ARPES
scans along a few lines parallel to the $\bar{\Gamma}-\bar{X}$ direction
of the surface Brillouin zone. Evolution of a tiny electron pocket
is visible around $k_{x}=0.33\,\protect\AA^{-1}$, as well as two
hole-like bands traversing the Fermi level around $k_{x}=0.22\,\protect\AA^{-1}$.
The spectra have been extracted from a map taken with a 0.5\textdegree{}
step in the polar angle. Excitation energy was $h\nu=68\,\mathrm{eV}$.
The $k_{y}=0$ spectrum is shown twice for the purpose of tracing
and labeling the bands, following their overall shape and ordering
from Figs. 6 and 7 in Ref. \onlinecite{Augustin2000}.}
\end{figure*}

By virtue of the photoexcitation process, the photon-energy dependency
measurements of a band dispersion can be translated into its $k_{z}$
dependence: $\hbar k_{z}=[2m_{e}(E_{\mathrm{kin}}\cos^{2}\vartheta-V_{o})]^{1/2}$.
In doing so, an assumption for the free-electron final state band
minimum $V_{o}$ has to be made, usually by taking into account the
band structure periodicity over a few Brillouin zones. We found that
the value of $-11.5\,\mathrm{eV}$, used in Ref. \onlinecite{Augustin2000},
is a reasonably good estimate for the inner potential $V_{o}$ and
have used it in our calculations.

The Brillouin zone size in the $\Gamma-Z$ ($k_{z}$) direction perpendicular
to WTe\textsubscript{2} layers is only $0.446\,\AA^{-1}$, and the
photoemission probes the higher order zones (in the extended scheme)
even with the lowest excitation energies. In our case this was the
sixth zone for 21~eV photons, and the ninth and tenth zone for the
40--70~eV range.

The band dispersion spectrum along the $k_{x}$ (chain direction)
at $k_{y}=0$ in Fig. \ref{fig:kz}(a) has been chosen from the $hv$
dependence map as it shows most features at the Fermi level. Two bands
found at $k_{x}$ of $0.28\,\AA^{-1}$ and $0.36\,\AA^{-1}$, are
strikingly flat along $k_{z}$, exhibiting localization within the
WTe\textsubscript{2} layers. The intensity arising from the band
centered around $k_{x}=0.16\,\AA^{-1}$ has only a limited $k_{z}$
extent around $\Gamma$ and reaches to about one half of the bulk
Brillouin zone. As localized states often exhibit large ARPES intensity
variations with the photon energy used, it is not clear if this represents
the actual $k_{z}$ range of the state or if it is an effect of varying
optical transition-probabilities in photoemission: the same band reappears
at much lower intensity around $k_{z}=4.47\,\AA^{-1}$ ($\Gamma_{10}$),
and two others can be identified, following the arrows in Fig. \ref{fig:kz},
with notable variation in intensity.

As all the bands crossing the Fermi level are also visible with 21~eV
photon energy we were able to map them in higher resolution. The ARPES
intensity maps shown in Fig. \ref{fig:FS20K} represent constant energy
contours of the states at and slightly below the Fermi level. The
states appear as two distinct pockets. The pocket at $0.33\,\AA^{-1}$
decreases in size as we go lower in energy, indicating that it is
the very bottom of a band. The opposite happens with the pocket around
$0.16\,\AA^{-1}$, and its area doubles going from the Fermi level
to just 60~meV below. The former thus represents an electron-like
(\emph{e}) pocket, and the latter a hole-like (\emph{h}) one. The
size of the pockets at the Fermi level is almost exactly the same,
$0.018\,\AA^{-2}$. Taking into account spin degeneracy and the fact
that there are two such pockets in the surface Brillouin zone, it
follows that the carrier concentration in the electron pockets is
about $1.8\cdot10^{13}\,\mathrm{cm}^{-2}$ and is nearly perfectly
compensated by the same concentration of holes. This is reminiscent
of the compensation in Mg, Zn, and Bi, all of which exhibit large
magnetoresistance, although with saturation at high field \cite{Stark1962,Falicov1966,Pippard1989,Behnia2009}.
When the Fermi surfaces for both electrons and holes are closed, i.e.
not connected across the borders of the Brillouin zone, the compensation
should lead to a quadratic rise of magnetoresistance in high magnetic
fields (see pp. 104--106 in \cite{Pippard1989}).  And indeed, a quadratic
rise was found for WTe\textsubscript{2} in Ref. \onlinecite{Ali2014}
for the fields between 1 and 60~T. This well balanced compensation
is likely the cause of the high saturation value of magnetoresistance:
it is reached in the magnetic fields that scale as the reciprocal
of the difference of charge densities, while its magnitude scales
as the square of that value. Thus, the better the compensation, the
higher the magnetoresistance limit (see pp. 28--31 in \cite{Pippard1989}).
Unlike any other known system at this time, the magnetoresistance
in WTe\textsubscript{2} does not appear to saturate even at the fields
of 60~T (in Bi, for example, it saturates by 40~T \cite{Behnia2009}).
It is important to note that the closeness of the two pockets as well
as their overall smallness may play an important role in the magnetoresistance:
the former through an increased scattering between the neighboring
orbits and the latter through the relation of the time needed to complete
an orbit to the mean scattering time. The extreme similarity of the
pockets and the lack of deviation from a quadratic dependence of the
magnetoresistance to very high fields implies that WTe\textsubscript{2}
may be the first example of a perfectly compensated semimetal \cite{Ali2014}.

A frieze of high resolution ARPES spectra taken along the chain direction,
Fig. \ref{fig:Dispersion_maps}, details the evolution of the electron
pocket (\emph{e}) as $k_{y}$ is swept across $\bar{\Gamma}$: the
whole pocket is accommodated within $\pm0.075\,\AA^{1}$ in both directions,
reaches down to 80~meV below the Fermi level, and approaches the
elbow of the rightmost band to within \textasciitilde{}20~meV. Exploiting
the similarities in the shape and ordering of the measured bands to
the calculated band structure, the bands along $\bar{\Gamma}$-$\bar{X}$
in Fig. \ref{fig:Dispersion_maps} have been labeled as in Ref. \onlinecite{Augustin2000}.
The electron pocket is likely formed by an avoided hybridization of
a tungsten 5\emph{d} band (I) and a tellurium 5\emph{p} band (III).
This band, while returning from above the Fermi level, and the band
originating from tungsten orbitals (V), make the hole pocket \cite{Augustin2000}.
To add to the complexity of the band structure, a linear band (VI)
dispersing at $v=5~\mathrm{eV}\AA$ passes right through the hole
pocket; its impact on the transport in WTe\textsubscript{2} is currently
unknown, but probably just adds a constant contribution independent
of the temperature.

Although the general shape of the density functional theory derived
bands in Refs. \onlinecite{Augustin2000} and \onlinecite{Ali2014}
corresponds to what has been measured here, quantitative agreement
is far from satisfying. As an example: the separation between the
elbow shaped band (III$\rightarrow$V) and the electron pocket (I$\rightarrow$III)
appears exaggerated in the calculations both in momentum and energy;
the fast dispersing band (VI) that is crossing the Fermi level at
$0.2\,\AA^{-1}$ is present in \cite{Augustin2000} but is missing
or appears highly hybridized in \cite{Ali2014}; most calculated bands
turn out to be dilated in momentum, and for some, considerable shifts
in energy can be noted. This calls for a better theoretical modeling
of the system.

\begin{figure}
\includegraphics[width=88mm]{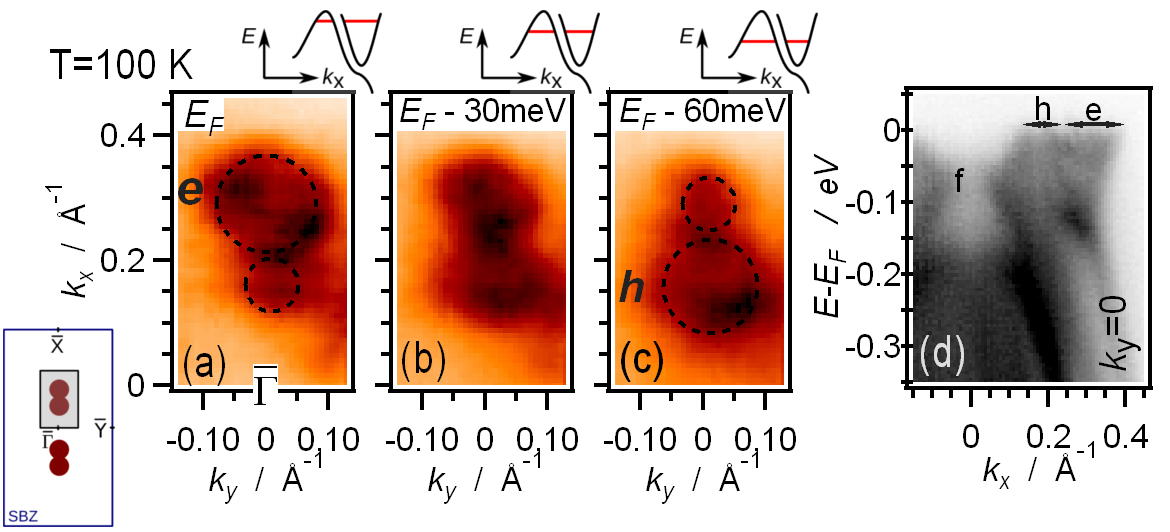}\protect\caption{\label{fig:FS100K} (a)--(c) Constant energy contours of the electron
(\emph{e}) and hole (\emph{h}) pockets in WTe\protect\textsubscript{2}
band structure close to the Fermi level, $E_{F}$, taken at T=100~K.
The mapping has been carried out by polar rotations of 0.25\textdegree .
(d) Low energy band dispersion acquired along the $\bar{\Gamma}-\bar{X}$
direction at T=100~K. }
\end{figure}

The relative increase of magnetoresistance with the magnetic field
in WTe\textsubscript{2} weakens with increasing temperature. Our
data show a dramatic change of the Fermi surface as the temperature
is raised from 20~K to 100~K. A constant energy map taken at 100~K,
Fig. \ref{fig:FS100K}(a), shows at the Fermi level a large electron
pocket centered at $k_{x}=0.30\,\AA^{-1}$ and a tiny hole pocket
around $k_{x}=0.16\,\AA^{-1}$. The temperature increase caused the
bands forming the pockets to become electron doped by about 30~meV,
as the contour with largely compensated electrons and holes that represented
the Fermi surface at 20~K is now found at a 30~meV lower energy.
Assuming a continuous evolution as the temperature is turned down,
a predominantly electron-like Fermi surface transforms into one with
perfectly compensated electrons and holes. Were the carrier compensation
the only mechanism, the upsurge of the magnetoresistance in WTe\textsubscript{2}
would indicate the balance being reached in the limit of 0~K. While
this can be considered the origin of the weakening of the effect with
increasing temperatures, a competing mechanism might be responsible
for its complete diminishing \cite{Ali2014} above 150~K.

A band centered at $\bar{\Gamma}$, marked by \emph{f} in Fig. \ref{fig:Dispersion_maps},
appears almost flat just about 65~meV below the Fermi level at 20~K,
and does not change position when the temperature is raised to 100~K,
Fig. \ref{fig:FS100K}(d). Its contour is also visible in the constant
energy map at -100~meV (shown in Fig. \ref{fig:twoSBZ}(b)). While
the band may be inert in transport at very low temperatures, Fermi-Dirac
statistics works in favor of its conduction when the material is warmed:
at 160~K, 1\% of the electrons at this single energy level are thermally
excited leaving conductive holes in the band, while this proportion
at 80~K is only 0.01\%. Combined with a high density of states related
to its flatness, it is likely that at higher temperatures this band
takes conduction over the bands that cross the Fermi level, and that
the proximity of this band to the Fermi level and the Landau quantization
of its orbits is responsible for the \emph{turn on} temperature behavior
found in Ref. \onlinecite{Ali2014}.

We have thus shown that tiny electron and hole pockets of equal size
are the electronic basis for the unusual transport properties of WTe\textsubscript{2}
at low temperatures. Extremely large magnetoresistance emerges from
the resulting charge compensation, which has been shown to be temperature
dependent, apparently approaching perfect compensation in the limit
of zero temperature. The pronounced anisotropy of the magnetoresistance
is associated with the uniaxial character of the Fermi surface and
proximity of the electron and hole pockets in momentum space along
the direction of the metallic chains in this quasi one-dimensional
material. A flat band lying below the Fermi level has been recognized
as the source of the \emph{turn on} temperature behavior of the magnetoresistance.
Band dispersion measurements uncovered a complex structure in the
vicinity of the Fermi level that calls for a more accurate band structure
calculation and, based on it, a quantitative theoretical modeling
of the magnetoresistance. 
\begin{acknowledgments}
This work was supported by the US Department of Energy, Office of
Basic Energy Sciences, contracts No. DE-AC02-98CH10886 and DE-AC02-05CH11231,
and ARO MURI program, grant W911NF-12-1-0461. ALS is operated by the
US DOE under contract No. DE-AC03-76SF00098. \end{acknowledgments}

\end{document}